\documentclass{article}
\usepackage[utf8]{inputenc}
\usepackage[T2A]{fontenc}
\usepackage[english]{babel}
\inputencoding{utf8}

\usepackage{amsmath,amssymb,amsthm}

\usepackage[pdfpagemode=UseNone,bookmarks=false,pdfpagelayout=SinglePage,pdfstartview={XYZ null null 1},
pdfhighlight=/P,colorlinks=true,linkcolor=blue,citecolor=blue,urlcolor=blue]{hyperref}

\advance\textwidth 20mm  \advance\oddsidemargin -10mm
\advance\textheight 20mm \advance\topmargin -15mm

\theoremstyle{plain}

\theoremstyle{definition}

\title{Nonabelian ${\mathfrak so}_3$ Euler top}

\date{22 December 2020}

\author{V.~Sokolov\thanks{L.D.~Landau Institute for Theoretical Physics, Chernogolovka, Russian Federation.}\,$^*$\thanks{Federal University of ABC, Santo Andr\'e, Sao Paulo, Brazil. E-mail: vsokolov@landau.ac.ru}}

\begin{document}
\maketitle

\begin{abstract}
Using the nonabilinization procedure, we find an integrable matrix version of the Euler top on $\mathfrak{so}_3$.  
\end{abstract}

\medskip
\medskip

\noindent{\small Keywords:  Euler equation, nonabelian system, Lax representation, first integral, symmetry

\medskip
\medskip

%-------------------------------------------------------------------------------
 
In the papers \cite{SW2, Adler_Sokolov_2020} an approach for constructing integrable noncommutative generalizations of a given polynomial integrable system was proposed.  It turns out that requiring the noncommutative generalization not only for the system itself but also for all its first integrals (conservation laws for PDEs) and infinitesimal symmetries, one can easily construct all such generalizations.  In this note we apply this way for finding noncommutative generalizations of the Euler top. Some other generalizations, where a part of first integrals and/or symmetries ``disappear'', are known. 

Consider the system of ODEs 
\begin{equation}\label{Sc}
 u'= z_1\, v w, \qquad v'= z_2\, u w, \qquad w'= z_3\, u v, \qquad  z_i\in \mathbb C, \quad z_i\ne 0,
\end{equation}
where $'$ means the derivative with respect to $t$. In the case $z_1=\frac{B-C}{A},\,\, z_2=\frac{C-A}{B},\,\,z_3=\frac{A-B}{C}\,\,$, where  $A,B,C$ are axis of the ellipsoid of inertia, equations \eqref{Sc} describe the Euler top. System \eqref{Sc} possesses the first integrals 
$I_1 = z_3 u^2 - z_1 w^2, \quad I_2=z_3 v^2-x_2 w^2.$ For any $i,j$ the system 
\begin{equation}\label{Sym}
 u_{\tau}= z_1\, v w \,\,I_1^i I_2^j, \qquad v_{\tau}= z_2\, u w \,\,I_1^i I_2^j, \qquad w_{\tau}= z_3\, u v \,\,I_1^i I_2^j  
\end{equation}
is an infinitesimal symmetry for \eqref{Sc}.

It is easy to see that the parameters $z_i$ can be always reduced to one by some (complex) scalling of the independent and dependent variables.  Everywhere below, we assume that $z_1=z_2=z_3=1.$\footnote{For some purposes the normalization $z_1=-1, \,\, z_2=z_3=1$ seems to be more reasonable. In particular, in this case the trajectories are compact.}

Our goal is to generalize system \eqref{Sc} to the case when the unknown variables become matrices of arbitrary size $n\times n.$ Assuming that the right hand sides are still homogeneous quadratic polynomials and that the system coincides with \eqref{Sc} for $n=1$, we arrive at the following ansatz with unknown constant coefficients:
\begin{equation}\label{anzats}
\begin{cases}
  u' = k_1 v w + (1-k_1) w v + c^1_{12} [v, w] + c^1_{23} [w,u] +c^1_{31} [u, v] , \\[1.5mm]
  v' = k_2 w u + (1-k_2) u w + c^2_{12} [w, u] + c^2_{23} [u, v] +c^2_{31} [v, w] \\[1.5mm]
	w' = k_3 u v + (1-k_3) v u + c^3_{12} [u, v] + c^3_{23} [v,w] +c^3_{31} [w, u].
\end{cases}
\end{equation}

To begin with, we require that for each integral $R_{ij}=I_1^i I_2^j$ of system \eqref{Sc} there exists an integral of system 
\eqref{anzats} of the form ${\it trace}\, P_{ij}$ (\cite[chapter 6.1]{Sokolov}), where $P_{ij}$ is a matrix polynomial, that coincides with $R_{ij}$ if $n=1$. We will call the integral ${\it trace}\, P_{ij}$  {\it nonabelinization} of the integral $R_{ij}$. 

{\bf Proposition 1.}  If there exist nonabelinizations of all integrals $R_{ij}$ with $i+j\le 3$,  then  system \eqref{anzats} has the form   
\begin{equation}\label{anzatsRed}
\begin{cases}
\displaystyle   u' = \frac{1}{2} (v w + w v) + X [u, v] + Z [u,w ], \\[2mm]
\displaystyle   v' = \frac{1}{2} (w u + u w) + Y [v, u] + Z [v, w], \\[2mm]
\displaystyle 	w' = \frac{1}{2} (u v + v u) + X [w, v] + Y [w, u],
\end{cases}
\end{equation}
where $X,Y,Z$ are arbitrary constant parameters.

{\bf Remark 1.} The permutation of the variables $ u, v, w $ leads to the corresponding permutation of the coefficients $ X, Y, Z $. In addition, one can change the signs of several coefficients. For example, the transformation $ u \to -u, \, t \to -t $ replaces $ Y \to -Y $.

For any parameters $X,Y,Z,$ system \eqref{anzatsRed} possesses a Lax representation   $L_t = [A,\,L]$ in the Lie algebra  
$$
{\cal G}=\left\{\begin{pmatrix}
   U & V\\[1mm]
  -V & U
  \end{pmatrix}, \qquad U,V\in {\mathbb H} \right\}
$$
of matrices over the skew-field of quaternions   (cf. \cite{Kimura}). It is assumed that quaternions commute with nonabelian variables $ u, v, w .$

The matrices $L$ and $A$ have the form  
$$
L = \begin{pmatrix}
   L_1 &  L_2\\[1mm]
  -L_2 & L_1
  \end{pmatrix}, \qquad A = \begin{pmatrix}
   A_1 & A_2\\[1mm]
  -A_2 & A_1
  \end{pmatrix},
$$ 
where
$$
L_1=2 (\nu - \mu) {\rm Det} (S) \, u, \qquad L_2  = \langle P, \Omega \rangle \,v + \langle Q, \Omega \rangle \, w, $$$$ A_1 = (-Y \, u-X\,v - Z\,w)\,{\bf 1} + \sigma L_1, \qquad A_2 = \mu \langle P, \Omega \rangle \,v + \nu\,\langle Q, \Omega \rangle \, w.
$$
Here $\sigma, \mu, \nu$ are arbitrary pairwise distinct parameters, $\Omega = ({\bf i},{\bf j},{\bf k})$, $P$ и $Q$ are 3--dimension vectors such that
$$
\langle P, Q \rangle = 0, \qquad \langle P, P \rangle = \frac{1}{4 (\sigma-\mu)(\nu-\mu)}, \qquad \langle Q, Q \rangle = \frac{1}{4 (\sigma-\nu)(\mu-\nu)}, 
$$
and $S$ is the matrix with rows $P,Q, \Omega$. Replacing the quaternion units ${\bf i},{\bf j},{\bf k}$ with the Pauli matrices 
$$
{\bf i}\to \begin{pmatrix}
   i & 0\\[1mm]
   0 & -i
  \end{pmatrix}, \qquad {\bf j}\to \begin{pmatrix}
   0 & 1\\[1mm]
   -1 & 0
  \end{pmatrix}, \qquad {\bf k}\to \begin{pmatrix}
   0 & i\\[1mm]
   i & 0
  \end{pmatrix},
$$
we obtain a Lax pair in $4\times 4$-matrices. This Lax pair depends on one essential parameter
$\displaystyle \kappa=\frac{(\sigma-\mu)(\nu-\mu)}{(\sigma-\nu)(\mu-\nu)}.$
 Other parameters  can be removed by means of a conjugation by a quaternion, by a shift $A \to A + {\rm const}\, L$ and a scalling  $t \to \tau^{-1} t, \,u \to \tau u,\,v \to \tau v,\,w \to \tau w$. For example, one may set $\sigma=0,\,\, \mu=1.$

Probably, the obtained Lax pair allows an algebraic $R$-matrix interpretation or a description in terms of a decomposition of the corresponding loop algebra into a sum of two subalgebras not for any set of parameters $X,Y,Z$.

{\bf Conjecture.} In the case of system \eqref{anzatsRed} all integrals $R_{ij}$ permit a nonabelinization. The corresponding nonabelian integrals are generated by the traces of powers of the operator $L$.

Apart from nonabelinization of the integrals $R_{ij}$, system \eqref{anzatsRed} has other integrals like ${\it trace}\, (u v w -u w v).$ In the commutative case these integrals vanish.

Systems \eqref{anzatsRed} with different parameters $X,Y,Z$ have non-isomorphic algebras of polynomial symmetries.

{\bf Remark 2.} System \eqref{anzatsRed} with $X=Y=Z=0$ can be obtained by a reduction from the Nahm equation  
$$
u' = [v, w], \qquad v' = [w, u], \qquad w' = [u, v].
$$
This system, like the Nahm equation itself, has first integrals, but does not have nonabelian polynomial symmetries. The latter fact was verified for symmetries of orders not higher than six.
 
 Now let us additionally require  the existence of nonabelinizations for all symmetries  \eqref{Sym} (\cite[chapter 6.1]{Sokolov}).
  
{\bf Proposition 2.} If system \eqref{anzatsRed} admits nonabelinization of symmetries \eqref{Sym} with $i=1, j=0,$ then it can be reduced to one of the following:
\begin{itemize}
\item[\bf 1.] \qquad $X=Y=Z=\frac{1}{2}$;
\item[\bf 2.] \qquad  $Y=Z=0, \,\, X = \frac{1}{2}$
\end{itemize}
by transformations from Remark 2. 

Apparently, system \eqref{anzatsRed} with these values of the parameters is ``more integrable'' than in the generic case. It would be interesting to understand how these sets of parameters are distinguished from the point of view of the Lax pair presented above.

%-------------------------------------------------------------------------------
\subsubsection*{Acknowledgements}

The author is grateful to M.~Dunajski  and V.~Rubtsov for useful discussions. This work was carried out under the State Assignment 0029-2021-0004 (Quantum field theory) of the Ministry of Science and Higher Education of the Russian Federation.
 
%-------------------------------------------------------------------------------

\end{document}